\documentclass[prd, twocolumn, showpacs, superscriptaddress, floatfix, nofootinbib]{revtex4}

\usepackage[dvips]{graphicx} \usepackage{graphicx} \usepackage{epsf} \usepackage{amsmath} \usepackage{amssymb}

\usepackage{graphicx} \usepackage{dcolumn} \usepackage{bm}

\def\be{\begin{equation}} \def\ee{\end{equation}} \def\bea{\begin{eqnarray}} \def\eea{\end{eqnarray}}

\usepackage{color}

\begin{document}

\title{Nonlinear Perturbations in a Variable Speed of Light Cosmology} \author{J. W. Moffat\\~\\ Perimeter Institute for Theoretical Physics, Waterloo,
Ontario N2L 2Y5, Canada\\ and\\ Department of Physics and Astronomy, University of Waterloo, Waterloo,\\ Ontario N2L 3G1, Canada}


\begin{abstract} A variable speed of light (VSL) cosmology is described in which the causal mechanism of generating primordial perturbations is achieved by varying the speed of light in a primordial epoch. This yields an alternative to inflation for explaining the formation of the cosmic microwave background
(CMB) and the large scale structure (LSS) of the universe. We make use of the $\delta{\cal N}$ formalism to identify signatures of primordial nonlinear
fluctuations, and this allows the VSL model to be distinguished from inflationary models. In particular, we find that the parameter $f_{\rm NL}=5$ in the
variable speed of light cosmology.  The value of the parameter $g_{\rm NL}$ evolves during the primordial era and shows a running behavior. \end{abstract}

\maketitle

\newcommand{\eq}[2]{\begin{equation}\label{#1}{#2}\end{equation}}

\section{Introduction}

Inflationary cosmology is successful in explaining the CMB observational data. However, it is not the unique paradigm of the very early universe. For
example, we have bouncing cosmology, string gas cosmology, and the variable speed of light (VSL) cosmology~\cite{Moffat}. We refer to
refs.~\cite{Novello:2008ra, Lehners:2008vx, Cai:2014bea, Battefeld:2014uga} for comprehensive reviews on bouncing cosmologies and
to~\cite{Brandenberger:2008nx, Battefeld:2005av} for reviews on string gas cosmology. It is necessary to obtain an observational signature that can
distinguish between the different competing models of the early universe. If primordial B-mode polarization gravitational waves can be
detected, then the measured ratio $r={\rm tensor/scalar}$ will help to distinguish between the models, for this requires a mechanism to produce a stretching of both scalar perturbation density and gravitational waves to superhorizon scales~\cite{Moffat}, so that they can be detected in the CMB. However, a measurement of the tensor mode tilt $n_t$ is needed to give a definitive test of the models. A joint analysis of the BICEP2/Keck Array observational data found strong evidence for foregound dust and no statistical significant evidence for gravitational wave tensor modes. An upper limit was obtained for the tensor-to-scalar mode ratio $r < 0.12$~\cite{BICEP2}

Another important test of competing models is the magnitude of nonlinear perturbations of the scalar and tensor power spectra. In the following, we will
calculate the magnitude of the nonlinear perturbations in a version of VSL~\cite{Moffat}, using the so-called $\delta{\cal N}$ formalism. We will find that
the nonlinear perturbations are of order unity, and the non-gaussian parameter $f_{\rm NL}$ and its running behavior $g_{\rm NL}$ are consistent with the Planck Mission data~\cite{PLANCK2015}.

\section{The VSL model}

The model is formulated using a Friedmann-Lema\^{i}tre-Robertson-Walker {FLRW} metric: \begin{equation} \label{FLRW}
ds^2=c^2dt^2-a^2\biggl[\frac{dr^2}{1-Kr^2}+r^2(d\theta^2+\sin^2\theta d\phi^2)\biggr], \end{equation} with $K=0,+1,-1$ for flat, closed and open models.
The metric has the group symmetry $O(3)\times R$ with a preferred proper comoving time $t$.

Setting the cosmological constant $\Lambda=0$, the variation of the action with respect to $g^{\mu\nu}$ yields the field equation~\cite{Moffat}:
\begin{equation} 
\label{Graveqs} G_{\mu\nu}=\frac{8\pi
G}{\Phi}T_{\mu\nu}+\frac{\nabla_\mu\partial_\nu\Phi}{\Phi}-g_{\mu\nu}\frac{\nabla^\alpha\nabla_\alpha\Phi}{\Phi}, \end{equation} 
where $G_{\mu\nu}=R_{\mu\nu}-\frac{1}{2}g_{\mu\nu}R$, $\nabla_\mu$ is the covariant derivative with respect to the metric $g_{\mu\nu}$ and $\Phi(x)=c^4(x)$. We also obtain the field equation for the scalar ``seed'' field $\phi$: \footnote [1] {From the action principle~\cite{Moffat}, we also have the field equation: 
\begin{equation} 
\label{psiDiffequation} \nabla_\mu (B^{\mu\nu})-\frac{\partial W(\psi_\mu)}{\partial\psi_\nu}=0, \end{equation} 
where $\psi_\mu$ is a vector field, $B_{\mu\nu}=\partial_\mu\psi_\nu-\partial_\nu\psi_\mu$ and $W(\psi_\mu)$ is a potential. For $\langle 0|\psi_\mu|0\rangle\neq 0$ the Lorentz group $SO(3,1)$ is spontaneously broken down to $O(3)$.} 
\begin{equation} 
\label{phiDiffequation} g^{\mu\nu}\nabla_\mu\nabla_\nu\phi+\frac{\partial
V(\phi)}{\partial\phi}=0, \end{equation} 
where $V(\phi)$ is a potential. In the action $S$, we have made the speed of light $c(x)$ a dynamical degree of
freedom. The energy-momentum tensor $T^{\mu\nu}$ satisfies the conservation law: 
\begin{equation} \nabla_\nu T^{\mu\nu}=0. 
\end{equation}

We postulate that fractions of seconds after the big bang $c$ undergoes a large superluminal increase $c\gg c_0$ where $c_0$ is the present measured speed of light. This solves the horizon and flatness problems ~\cite{Moffat}. Moreover, the primordial quantum scalar matter fluctuations and gravitational
wavelengths $\lambda_s$ and $\lambda_w$, respectively,  are stretched superhorizon for $c\gg c_0$ and frozen in as classical fluctuation imprints on the
CMB: \begin{equation} \lambda_s=\frac{c}{\nu_s},\quad \lambda_w=\frac{c}{\nu_w}, \end{equation} where $\nu_s$ and $\nu_w$ denote the constant values of the scalar and gravitational wave frequencies, respectively.

The Friedmann equations in our model are given by 
\begin{equation} \label{Friedmann} H^2+\frac{Kc^2}{a^2}=\frac{8\pi G\rho}{3}-4H\frac{\dot c}{c},
\end{equation} 
and 
\begin{equation} 
\label{Friedmann2} \frac{\ddot a}{a}=-\frac{4\pi G}{3}\biggl(\rho+3\frac{p}{c^2}\biggr)-2H\frac{\dot c}{c}-6\frac{{\dot
c}^2}{c^2}-2\frac{\ddot c}{c}, \end{equation} 
where $H={\dot a}/{a}$ and $\rho=\rho_M+\rho_\psi+\rho_\phi$.

The scalar field $\phi$ that describes the structure growth ``seed'' field satisfies the field equation: 
\begin{equation}
\ddot\phi+3H\dot\phi+c^2V(\phi)=0. 
\end{equation}
We assume that $V(\phi)=0$ and obtain the solution: 
\begin{equation} \label{dotphi}
\dot\phi=\sqrt{12}B\biggl(\frac{a_*}{a}\biggr)^3, \end{equation} 
where $B$ is a constant and $a_*$ is a reference value for $a$. For large $\dot\phi$ the kinetic contribution to $\rho_\phi\propto\frac{1}{2}(\dot\phi)^2 $ will dominate the matter densities $\rho_M$ and $\rho_\psi$ and the Friedmann equation (\ref{Friedmann}) becomes 
\begin{equation} H^2+\frac{Kc^2}{a^2}=\frac{1}{12}\dot\phi^2. \end{equation} 
Substituting the solution (\ref{dotphi}), we get
\begin{equation} H^2+\frac{Kc^2}{a^2}=B^2\biggl(\frac{a_*}{a}\biggr)^6. 
\end{equation} 
Because the field $\phi$ dominates in the early universe, we can
neglect the spatial curvature for $c=c_0$ and we obtain the approximate solution given by 
\begin{equation} \label{asolution} a(t)=a_*(3Bt)^{1/3},\quad H(t)=\frac{1}{3t}. 
\end{equation} 
The equation of state for a massless scalar field gives for the exponent $n$ in $a(t)\propto t^n$ the value $n=2/3(1+w)$,
so for $n=1/3$ we get $w=1$.  A calculation of the power spectra and spectral indices (tilts) for the scalar perturbation density fluctuations and gravitational wave tensor fluctuations yields: $n_s=0.96$ and the red tilt $n_t=- 0.04$~\cite{Moffat}. If we adopt the tensor/scalar ratio bound $r < 0.12$~\cite{BICEP2}, then we have $\vert r/n_t\vert < 3$.

\section{Non-gaussianities in the VSL cosmology}

In this section, we first briefly review the standard process of calculating nonlinear perturbations by virtue of the so-called $\delta{\cal N}$ formalism.
Within the context of the VSL cosmology, primordial fluctuations are seeded by the scalar field $\phi$ and, therefore, the mechanism of generating
primordial curvature perturbation in the VSL cosmology is analogues to the curvaton scenario~\cite{Lyth:2001nq, Mollerach:1989hu, Linde:1996gt,
Enqvist:2001zp}. Then, with the assumption that the end of the epoch $c\gg c_0$ when $c=c_0$ is a uniform total density slice, we generalize the
$\delta{\cal N}$ formalism~\cite{Salopek:1990jq, Sasaki:1995aw, Sasaki:1998ug, Wands:2000dp, Lyth:2004gb} to be applicable to our case. Afterwards, we
analyze the non-gaussianities of primordial curvature perturbations in VSL cosmology.

\subsection{The generalized $\delta{\cal N}$ Formalism}

The $\delta {\cal N}$ formalism is available upon two conditions: firstly, the universe is isotropic and homogenous at extremely large scales that can accommodate a large number of causally isolated regions; secondly, the perturbation is frozen after the horizon exit. These two conditions are easily met in inflationary cosmology. However, for alternatives to inflation, if one can argue based on them, the $\delta {\cal N}$ formalism is still available. The first condition is in fact related to the horizon problem and, hence can be satisfied in quite a number of alternative models. For the second condition, it is not satisfied in matter bounce cosmology and marginally satisfied in ekpyrotic cosmology. In VSL models, the universe is still expanding with an effective equation of state parameter $w=1$, and as a result, the dominant perturbation grows as a logarithmic function. In this case, the second condition is also satisfied and from this argument the $\delta {\cal N}$ formalism is available as well in VSL. 

On superhorizon scales, if we assume negligible interaction between primordial scalar field $\phi$ and other matter fields, the component curvature
perturbation on the uniform density slice can be identified as 
\begin{eqnarray} \label{delta_N}
 \zeta_i (x) = \delta{\cal N} (x) + \frac{1}{3} \int_{\bar\rho(t)}^{\rho_i(x)} \frac{d \tilde\rho_i}{\tilde\rho_i+P_i(\tilde\rho_i)},
\end{eqnarray} 
through the $\delta{\cal N}$ formalism, where the subscript $i$ represents the component matter field in the universe.

Having the component curvature perturbation in mind, we can calculate the curvature perturbation on the uniform density slice. In analogy with the curvaton mechanism, it is natural to choose this slice at the end moment of $c\gg c_0$ period when $c=c_0$. On the uniform density slice we have \begin{eqnarray}
 \rho_r+\rho_\phi = \bar\rho,
\end{eqnarray} 
where $\rho_r$ and $\rho_\phi$ represent the radiation and scalar field densities, respectively. For the primordial scalar field, the
background dynamics behaves as a stiff fluid with an effective equation of state $w_\phi = 1$. Then, following Eq. \eqref{delta_N}, we can
obtain~\cite{Cai:2010rt}: \begin{eqnarray}
 \rho_r = \bar\rho_r e^{4(\zeta_r-\zeta)},\quad\rho_\phi = \bar\rho_\phi e^{6(\zeta_\phi-\zeta)}.
\end{eqnarray} As a consequence, the curvature perturbation $\zeta$ can be derived on the uniform density slice through the relation: 
\begin{eqnarray}
\zeta = \tilde{r} \zeta_\phi , \end{eqnarray} where \begin{eqnarray} \tilde{r} = \frac{3\Omega_\phi}{2+\Omega_\phi}. \end{eqnarray} 
Here, we have introduced the transfer efficiency coefficient $\tilde{r}$ and $\Omega_\phi \equiv \bar\rho_\phi/\bar\rho_{tot}$ is the density parameter for the primordial scalar field. Note that, in a generic case, $\tilde{r} = 3(1+w_\phi)\Omega_\phi/[4+(3w_\phi-1)\Omega_\phi]$, and we have applied the relation $w_\phi =1$. In the VSL cosmology, we have $\Omega_\phi = 1$, for the scalar field dominates over other matter fields at the end of the primordial era. This implies that the transfer from the scalar field fluctuations into primordial curvature perturbations is very efficient and thus, 
$\tilde{r} = 1$.

Within the local ansatz of the curvature perturbation, one can expand its form order by order as follows, 
\begin{align}
 \zeta(x) &= \zeta_1(x) + \zeta_2(x) + \zeta_3(x) + O(\zeta_4) \nonumber\\
          &= \zeta_1(x) +\frac{3}{5}f_{\rm NL}\zeta_1^2(x) +\frac{9}{25}g_{\rm NL}\zeta_1^3(x) + O(\zeta_4),
\end{align} 
where $\zeta_1$ is the fluctuation of the Gaussian distribution, and $\zeta_n$ are the non-Gaussian fluctuations. From the above expansion, we
obtain 
\begin{eqnarray}\label{fg_NL}
 f_{\rm NL} = \frac{5}{3}\frac{\zeta_2}{\zeta_1^2} , \quad g_{\rm NL} = \frac{25}{9}\frac{\zeta_3}{\zeta_1^3},
\end{eqnarray} for the local type. To be general, one can relate the nonlinearity parameters to the bispectrum and the trispectrum via: \begin{align}
 &B({\bf k}_1, {\bf k}_2, {\bf k}_3) = \frac{6}{5} f_{\rm NL} [P(k_1)P(k_2)+2 {\rm perm}]~, \nonumber\\
 &T({\bf k}_1, {\bf k}_2, {\bf k}_3, {\bf k}_4) = \frac{54}{25} g_{\rm NL} [P(k_1)P(k_2)P(k_3)+3 {\rm perm}] \nonumber\\
 & ~~~~~~~~~~ +\tau_{\rm NL}[P(k_1)P(k_2)P(|{\bf k}_1+{\bf k}_3|) +11 {\rm perm}],
\end{align} 
where these spectra are associated with the correlation functions as follows: 
\begin{align}
 &\langle \zeta({\bf k}_1)\zeta({\bf k}_2) \rangle = (2\pi)^3 P(k_1) \delta^{(3)}(\sum_{a=1}^2{\bf k}_a) ~,\\
 &\langle \zeta({\bf k}_1)\zeta({\bf k}_2)\zeta({\bf k}_3) \rangle = (2\pi)^3 B({\bf k}_1, {\bf k}_2, {\bf k}_3) \delta^{(3)}(\sum_{a=1}^3{\bf k}_a) ~,
 \nonumber\\
 &\langle \zeta({\bf k}_1)\zeta({\bf k}_2)\zeta({\bf k}_3)\zeta({\bf k}_4) \rangle\nonumber\\
&~~~~~~~~~= (2\pi)^3 T({\bf k}_1, {\bf k}_2, {\bf k}_3, {\bf k}_4) \delta^{(3)}(\sum_{a=1}^4{\bf k}_a).\nonumber \end{align} 
Moreover, the two point correlation function is related to the regular power spectrum through the relation: 
\begin{eqnarray}
 {\cal P}_\zeta (k) = \frac{k^3}{2\pi^2} P(k).
\end{eqnarray}

\subsection{Primordial perturbations and non-Gaussianities in the VSL cosmology}

Following the discussion in the previous subsection, one can find that the dynamics of primordial curvature perturbation would be identified by the
variation of the generalized VSL e-folding number by making use of the local ansatz. To calculate such a variation, we need to know the evolution of the
background universe. The scalar field satisfies the Klein-Gordon equation: 
\begin{eqnarray}\label{KG_fast}
 \ddot\phi +3 H\dot\phi = 0.
\end{eqnarray} 
In this phase, $a\propto t^{1/3}$ and, therefore, the Hubble parameter is expressed as 
\begin{eqnarray} 
\label{H_fast}
 H(t) = \frac{1}{3t}.
\end{eqnarray}

According to the convention of ref.~\cite{Moffat}, the primordial scalar field $\phi$ is dimensionless. Substituting Eq. \eqref{H_fast} into the background
equation of motion \eqref{KG_fast} yields the following solution: \begin{eqnarray} \label{phigamma} \phi(t) = \phi_f + \gamma \ln\bigg(
\frac{t}{t_f}\bigg),\quad \dot\phi (t) = \frac{\gamma}{t}, \end{eqnarray} where $\phi_f$ denotes the value of the scalar field at the end moment of the
phase $t_f$ when $c=c_0$. The coefficient $\gamma$ is an integration constant that can be determined by the background Friedmann equation. Without loss of generality, we keep $\gamma$ as a free coefficient and see its effect on the variation of the effective e-folding number.

In our case, the VSL equivalent of the e-folding number is given by 
\begin{eqnarray}
 {\cal N} \equiv \int_t^{t_f} H dt = \frac{1}{3} \ln \bigg( \frac{t}{t_f} \bigg),
\end{eqnarray} 
and, correspondingly, the trajectories of the background dynamics can either be described by the phase space of $(\phi, \dot\phi)$ or that
of $({\cal N}, \gamma)$. In this regard, the method of calculating $\delta{\cal N}$ is similar to the analysis of~\cite{Chen:2013eea}.  However, the
effects of the background evolution of the VSL cosmology upon curvature perturbations are analogous to those occurring in the study of the matter bounce cosmology~\cite{Cai:2009fn, Cai:2011zx}, {\it where there is no quasi de Sitter expansion}. To be explicit, from now on we switch to the phase space of $(\phi, \gamma)$. Consequently, the VSL equivalent of the e-folding number takes the form: 
\begin{eqnarray}\label{N_efold}
 {\cal N} = {\cal N}(\phi, \lambda) = \frac{\phi-\phi_f}{3\gamma}.
\end{eqnarray}

By expanding the scalar field $\phi$, and determining the integration constant $\gamma$ via $\phi\rightarrow \phi+\delta\phi$ and $\gamma\rightarrow
\gamma+\delta\gamma$, we can determine the curvature perturbations order by order up to third order: 
\begin{align}\label{zeta_123}
 &\zeta_1 = {\cal N}_{,\phi}\delta\phi, \\
 &\zeta_2 = \frac{1}{2}{\cal N}_{,\phi\phi}\delta\phi^2 + {\cal N}_{,\phi\gamma}\delta\phi\gamma + \frac{1}{2}{\cal N}_{,\gamma\gamma}\delta\gamma^2,
 \nonumber\\
 &\zeta_3 = \frac{1}{6}{\cal N}_{,\phi\phi\phi}\delta\phi^3 + \frac{1}{3}{\cal N}_{,\phi\phi\gamma}\delta\phi^2\gamma
+ \frac{1}{3}{\cal N}_{,\phi\gamma\gamma}\delta\phi\gamma^2 \frac{1}{6}{\cal N}_{,\gamma\gamma\gamma}\delta\gamma^3.\nonumber \end{align} 
In the above expression, the subscript $_{,\phi}\equiv \partial/\partial\phi$ denotes the derivative with respect to $\phi$. After having obtained the expression ${\cal N}$ in \eqref{N_efold}, we can  explicitly determine the coefficients such as ${\cal N}_{,\phi}$. The above expression automatically includes the assumption that the field fluctuations are described before the primordial era by a highly Gaussian distribution.

From \eqref{phigamma} we get 
\begin{eqnarray}
 \gamma = \gamma(\phi, \dot\phi) = \frac{\phi-\phi_f}{{\cal W}\bigg(\frac{\phi-\phi_f}{t_f\dot\phi}\bigg)},
\end{eqnarray} 
where ${\cal W}$ is the Lambert function. As a consequence, we can find \begin{eqnarray}
 \delta\gamma \simeq \gamma_{,\phi}\delta\phi = \frac{\gamma}{\gamma+\phi-\phi_f}\delta\phi,
\end{eqnarray} where the contribution of $\delta\dot\phi$ is secondary at superhorizon scales. Inserting the above expression into Eq. \eqref{zeta_123}, we can derive 
\begin{align}
 &\zeta_1 = -\frac{1}{3(\gamma+\phi-\phi_f)} \delta\phi ~, \\
 &\zeta_2 = \frac{1}{3(\gamma+\phi-\phi_f)^2} \delta\phi^2 ~, \\
 &\zeta_3 = -\frac{(2\gamma-\phi+\phi_f)}{9\gamma(\gamma+\phi-\phi_f)^3} \delta\phi^3.
\end{align}

Recall that the nonlinearity parameters satisfy the relations in \eqref{fg_NL}. Consequently, these parameters can be identified in the VSL cosmology as
follows: 
\begin{align} \label{fg_NL_simple}
 &f_{\rm NL} = 5, \nonumber\\
 &g_{\rm NL} = \frac{25(2\gamma-\phi+\phi_f)}{3\gamma}.
\end{align} 
From Eq. \eqref{fg_NL_simple}, we observe that the nonlinearity parameter at second order $f_{\rm NL}$ is a positive constant of order unity
and thus is almost scale invariant.  However, the nonlinearity parameter at third order $g_{\rm NL}$ is a function of the scalar field during the
primordial phase and therefore implies a strong scale dependence.

\section{Conclusions and Discussion}

We have derived in our variable speed of light (VSL) model the primordial nonlinear fluctuations. We find that the parameter $f_{NL}=5$ which is compatible with the Planck2015 result: $f_{NL}=2.7\pm 5.8$~\cite{PLANCK2015}. The parameter $g_{NL}$ evolves during the primordial era and displays a running behavior. These result will allow the VSL model to be distinguished from inflationary models and other models such as the bouncing and string gas cosmology models.

\begin{acknowledgments} 

I thank Yi-Fu Cai, Robert Brandenberger and Viktor Toth for helpful and stimulating discussions. I thank Yi-Fu Cai for providing
calculations of the primordial nonlinear fluctuations. \end{acknowledgments}


\begin{thebibliography}{99}

\bibitem{Moffat}
  J. W. Moffat,
  ``Variable Speed of Light Cosmology, Primordial Fluctuations and Gravitational Waves,''
  arXiv:1404.5567 [astro-ph.CO].

\bibitem{Brandenberger:2011gk}
  R. H. Brandenberger,
  ``Introduction to Early Universe Cosmology,''
  PoS ICFI {\bf 2010}, 001 (2010), arXiv:1103.2271 [astro-ph.CO].

\bibitem{Novello:2008ra}
  M.~Novello and S.~E.~P.~Bergliaffa,
  ``Bouncing Cosmologies,''
  Phys.\ Rept.\  {\bf 463}, 127 (2008), arXiv:0802.1634 [astro-ph].

\bibitem{Lehners:2008vx}
  J. -L. Lehners,
  ``Ekpyrotic and Cyclic Cosmology,''
  Phys.\ Rept.\  {\bf 465}, 223 (2008), arXiv:0806.1245 [astro-ph].

\bibitem{Cai:2014bea}
  Y. -F. Cai,
  ``Exploring Bouncing Cosmologies with Cosmological Surveys,'' arXiv:1405.1369 [hep-th].

\bibitem{Battefeld:2014uga}
  D. Battefeld and P. Peter,
  ``A Critical Review of Classical Bouncing Cosmologies,''
  arXiv:1406.2790 [astro-ph.CO].

  \bibitem{Brandenberger:2008nx}
R. H. Brandenberger,
  ``String Gas Cosmology,''
  arXiv:0808.0746 [hep-th].

\bibitem{Battefeld:2005av}
  T. Battefeld and S. Watson,
  ``String gas cosmology,''
  Rev.\ Mod.\ Phys.\  {\bf 78}, 435 (2006), arXiv:hep-th/0510022.

\bibitem{BICEP2}
  P. A. R. Ade et al.,
  ``A Joint Analysis of BICEP2/Keck Array and Planck Data,'' Phys.\ Rev.\ Lett. {\bf 114}, 101301 (2015), arXiv:1502.00612 [astro-ph.CO].

\bibitem{PLANCK2015}
  P. A. R. Ade et al.,
  ``Planck 2015 results. XVII. Constraints on primordial non-Gaussianity,'' arXiv:1502.01592 [astro-ph.CO].

\bibitem{Lyth:2001nq}
  D. H. Lyth and D. Wands,
  ``Generating the curvature perturbation without an inflaton,''
  Phys.\ Lett.\ B {\bf 524}, 5 (2002), arXiv:hep-ph/0110002.

\bibitem{Mollerach:1989hu}
  S. Mollerach,
  ``Isocurvature Baryon Perturbations and Inflation,''
  Phys.\ Rev.\ D {\bf 42}, 313 (1990).

\bibitem{Linde:1996gt}
  A. D. Linde and V. F. Mukhanov,
  ``Nongaussian isocurvature perturbations from inflation,''
  Phys.\ Rev.\ D {\bf 56}, 535 (1997), arXiv:astro-ph/9610219.

\bibitem{Enqvist:2001zp}
  K. Enqvist and M. S. Sloth,
  ``Adiabatic CMB perturbations in pre - big bang string cosmology,''
  Nucl.\ Phys.\ B {\bf 626}, 395 (2002), arXiv:hep-ph/0109214.

\bibitem{Salopek:1990jq}
  D. S. Salopek and J. R. Bond,
  ``Nonlinear evolution of long wavelength metric fluctuations in inflationary models,''
  Phys.\ Rev.\ D {\bf 42}, 3936 (1990).

\bibitem{Sasaki:1995aw}
  M. Sasaki and E. D. Stewart,
  ``A General analytic formula for the spectral index of the density perturbations produced during inflation,''
  Prog.\ Theor.\ Phys.\  {\bf 95}, 71 (1996), arXiv:astro-ph/9507001.

\bibitem{Sasaki:1998ug}
  M. Sasaki and T. Tanaka,
  ``Superhorizon scale dynamics of multiscalar inflation,''
  Prog.\ Theor.\ Phys.\  {\bf 99}, 763 (1998), arXiv:gr-qc/9801017.

\bibitem{Wands:2000dp}
  D. Wands, K. A. Malik, D. H. Lyth and A. R. Liddle,
  ``A New approach to the evolution of cosmological perturbations on large scales,''
  Phys.\ Rev.\ D {\bf 62}, 043527 (2000), arXiv:astro-ph/0003278.

\bibitem{Lyth:2004gb}
  D. H. Lyth, K. A. Malik and M. Sasaki,
  ``A General proof of the conservation of the curvature perturbation,''
  JCAP {\bf 0505}, 004 (2005), arXiv:astro-ph/0411220.

\bibitem{Cai:2010rt}
  Y. -F. Cai and Y. Wang,
  ``Large Nonlocal Non-Gaussianity from a Curvaton Brane,''
  Phys.\ Rev.\ D {\bf 82}, 123501 (2010), arXiv:1005.0127 [hep-th].

\bibitem{Chen:2013eea}
  X. Chen, H. Firouzjahi, E. Komatsu, M. H. Namjoo and M. Sasaki,
  ``In-in and $\delta N$ calculations of the bispectrum from non-attractor single-field inflation,''
  JCAP {\bf 1312}, 039 (2013), arXiv:1308.5341 [astro-ph.CO].

\bibitem{Cai:2009fn}
  Y. -F. Cai, W. Xue, R. Brandenberger and X. Zhang,
  ``Non-Gaussianity in a Matter Bounce,''
  JCAP {\bf 0905}, 011 (2009), arXiv:0903.0631 [astro-ph.CO].

\bibitem{Cai:2011zx}
  Y. -F. Cai, R. Brandenberger and X. Zhang,
  ``The Matter Bounce Curvaton Scenario,''
  JCAP {\bf 1103}, 003 (2011), arXiv:1101.0822 [hep-th].

\end{thebibliography}
\end{document}